\documentclass[11pt,oneside,letterpaper]{article}

\usepackage[colorlinks=true,citecolor=darkgreen,linkcolor=purple,urlcolor=purple]{hyperref}
\usepackage{amssymb}
\usepackage{amsmath}
\usepackage[dvips]{graphicx}
\usepackage{setspace}
\usepackage{fancyhdr}
\usepackage{xcolor}
\usepackage{ifpdf}
\usepackage{graphicx}
\usepackage{rotating}
\usepackage{comment}

\definecolor{darkgreen}{rgb}{0,0.5,0}
\definecolor{darkblue}{rgb}{0,0,0.6}
\definecolor{purple}{rgb}{0.4,.2,0.7}
\definecolor{awesome}{rgb}{1.0, 0.13, 0.32}

\newcommand{\nn}{\nonumber}

\newcommand{\nin}{\noindent}
\newcommand{\bi}{\begin{itemize}}
\newcommand{\ei}{\end{itemize}}
\newcommand{\bea}{\begin{eqnarray}}
\newcommand{\eea}{\end{eqnarray}}
\newcommand{\be}{\begin{equation}}
\newcommand{\ee}{\end{equation}}
\newcommand{\non}{\nonumber}

 \newcommand{\red}[1]{{\color{red}#1}}
 \newcommand{\blue}[1]{{\color{blue}#1}}

\numberwithin{equation}{section} 

\begin{document}

\vspace*{1.5cm}
\begin{center}
{ \LARGE \textsc{ Solvable Quantum Grassmann Matrices}}
\\ \vspace*{1.7cm}
Dionysios Anninos$^1$ and Guillermo A. Silva$^{1,2}$
\\
\vspace*{1.7cm}
$^1$ {\it Institute for Advanced Study, 1 Einstein Drive, Princeton, NJ, 08540} \\
$^2$  {\it Instituto de F\'\i sica de La Plata - CONICET \& \\
Departamento de F\'\i sica, Universidad Nacional de La Plata} \\
1900, La Plata, Argentina

\vspace*{0.6cm}
\end{center}
\vspace*{1.5cm}
\begin{abstract}
\noindent
 
We explore systems with a large number of fermionic degrees of freedom subject to non-local  interactions. We study both vector and matrix-like models with quartic interactions. The exact thermal partition function is expressed in terms of an ordinary bosonic integral, which has an eigenvalue repulsion term in the matrix case. We calculate real time correlations at finite temperature and analyze the thermal phase structure. When possible, calculations are performed in both the original Hilbert space as well as the bosonic picture, and the exact map between the two is explained. At large $N$, there is a phase transition to a highly entropic high temperature phase from a low temperature low entropy phase. Thermal two-point functions decay in time in the high temperature phase. 

\end{abstract}

\newpage
\setcounter{page}{1}
\pagenumbering{arabic}

\setcounter{tocdepth}{2}

\onehalfspacing

\section{Introduction}

In this paper we are interested in the physics of a large number of non-locally interacting fermionic degrees of freedom. We will study quantum mechanical fermions with a vector-like index structure as well as a matrix-like index structure. Part of the motivation comes from recent investigation of similar systems \cite{Sachdev:1992fk,Kitaev,Sachdev:2015efa,Polchinski:2016xgd,MaldaStan,Anninos:2016szt,Jevicki:2016bwu,Cotler:2016fpe} which may play a useful role in understanding certain problems in black hole physics. Another motivation comes from recent investigations on the emergence of bosonic matrix models from discrete systems \cite{parisi,Anninos:2014ffa,Anninos:2015eji,sean,berenstein}. Moreover, we are interested in how the original fermionic degrees of freedom and Hilbert space might be encoded (if at all) in those of the bosonic matrix.\footnote{In a very broad sense, this perhaps is somewhat analogous to how the `deconfined gluons' of $\mathcal{N}=4$ super Yang-Mills at large $N$ are encoded in the bulk gravitational/string degrees of freedom propagating in AdS$_5\times S^5$.} Our goal is to study tractable systems with rich interactions. Unlike the purely fermionic models of \cite{Kitaev}, our models do not have any quenched disorder (see also \cite{Witten:2016iux,Gurau:2016lzk,Klebanov:2016xxf,Nishinaka:2016nxg}). Instead, their complexity stems from the matrix-like interactions. Given a matrix structure, we might hope, eventually, to relate our systems to some gravitational description for a new class of models. 

Here, we take some small steps in this direction by carefully analyzing and solving such systems with quartic interactions. Our approach is to express the systems in terms of auxiliary bosonic variables allowing us to use standard large $N$ techniques.  When possible, we carry forward the analysis in both the fermionic Hilbert space picture as well as the bosonic path integral picture and map physical questions from one picture to another. The bosonic systems exhibit a $(0+1)$-dimensional emergent gauge field and the corresponding (broken) gauge symmetry is crucial in our ability to solve the systems. Part of our treatment is in some regard a $(0+1)$-dimensional analogue of the analysis in \cite{Polyakov:1983tt,Polyakov:1984et}. Fermionic correlation functions are related to the calculation of certain Wilson line operators of the gauge field. At the end, for both matrix and vector models, we are able to express the exact finite $N$ thermal partition function as an ordinary integral. In the matrix case, this integral is a matrix-like integral with a modified Vandermonde term that consists of the $\sinh$ or $\sin$ of the difference in eigenvalues rather than the difference eigenvalues themselves. Such eigenvalue integrals appear in the study of unitary random matrix models as well as Chern-Simons theories. At large $N$ the systems exhibit a non-trivial phase structure which is naturally characterized, in the matrix case, by the connectivity of the eigenvalue distribution, somewhat similar to the situation encountered in \cite{sean}. We note that ordinary fermionic matrix integrals was considered in \cite{Makeenko:1993jg,Semenoff:1996vm,Paniak:2000zy}.

The structure of the paper goes as follows: we begin by analyzing the vector model in the early sections. We show how the fermionic thermal partition function, initially expressed as a path integral over Grassmann valued functions of Euclidean time, can be expressed in terms of an ordinary bosonic integral over a single real variable. We analyze both the thermal phase structure and fermionic correlation functions. The latter are expressed in terms of the expectation values of non-local in time variables, resembling Wilson line operators of an emergent gauge field. We show, at large $N$, a transition from a low entropy phase to one with entropy extensive in $N$. These results are generalized to the matrix case in the latter sections, where the structure is significantly more intricate. The thermal partition function now reduces to a matrix integral, which as mentioned, contains a modified Vandermonde interaction among the eigenvalues. We end discussing the thermal phase structure and thermal correlation functions of the matrix model at large $N$. At large $N$, we find that the real time two-point function decays significantly faster in the matrix case than the vector case. In appendices \ref{genpot} and \ref{gammaneg} we discuss certain generalizations of the models studied in the main body.

\section{Fermionic vector model}

In this section we consider $N$ complex Grassmann degrees of freedom $\{\psi^I,\bar{\psi}^I\}$ interacting via a quartic interaction. The index $I=1,2,\ldots,N$ is a $U(N)$ index, under which the $\psi^I$ transform in the fundamental and the $\bar{\psi}^I$ in the anti-fundamental representation. 

The thermal partition function of our model is given by:
\be
Z[\beta]= {\cal N}\int \mathcal{D} \psi^I(\tau) \mathcal{D} \bar\psi^I(\tau)  \,  e^{-\oint d\tau \,  \left({\bar{\psi}}^I(\tau) \dot{{\psi}}^I(\tau) -\frac1{4N\gamma}   \left(\bar{\psi}^I(\tau) {\psi}^I (\tau)\right)^2 \right)}~.
\label{Zpsi}
\ee
We have a periodic Euclidean time coordinate $\tau \sim \tau + \beta$, such that our integration variables obey $\psi^I(\tau+\beta) = -\psi^I(\tau)$. The fermion variables are dimensionless and the parameter $\gamma$ is a positive number with units of $\tau$. (In appendix \ref{gammaneg} we carry over our results for the vector model to the case with negative $\gamma$.) 
Using a Hubbard-Stratanovich transformation the Euclidean action can be written as
\begin{equation}
\label{faction}
S_E[\psi^I(\tau),\bar\psi^I(\tau),\lambda(\tau)] =  \oint d\tau \,  \left({\bar{\psi}}^I(\tau) \dot{{\psi}}^I(\tau) + {\lambda}(\tau)   \bar{\psi}^I(\tau) {\psi}^I(\tau) + N\gamma \lambda(\tau)^2 \right)~,
\end{equation}
with $\lambda(\tau)$ a periodic function of $\tau$ with units $\tau^{-1}$. The only dimensionless quantity, other than $N$, is $\gamma/\beta$. It grows with increasing temperature. We set $\beta = 1$ unless otherwise stated.

\subsection{Exact evaluation of path integral}

In this subsection we evaluate the fermionic path integral exactly. Going back to (\ref{faction}), evaluation of the fermionic path integral yields:
\begin{equation}
\label{bosonic}
Z = {\cal N}\int \mathcal{D}\lambda(\tau) \, {\det}^N \left[ \partial_\tau + \lambda(\tau) \right] e^{- N\gamma  \oint d\tau \lambda(\tau)^2}~.
\end{equation}
{The functional determinant has a local invariance \cite{Anninos:2015eji}:
\begin{equation}
\tau \to f(\tau)~, \quad \lambda(\tau) \to  \lambda(\tau)/f'(\tau)~.
\label{diffeo}
\end{equation}
This can be viewed as a time-reparameterization of an einbein $\lambda(\tau)$, and it is an exact symmetry of a system of non-interacting fermions. Due to the above invariance, the functional determinant will only depend on the Matsubara zero-mode of $\lambda(\tau)$, namely $\lambda_0 \equiv \oint d\tau\lambda(\tau)$. Let us see this perturbatively. Working in Fourier space, and taking into account that Fermions are anti-periodic along the thermal circle, we are interested in the object
\begin{align}
\log\det[\partial_\tau + \lambda(\tau) ] = \text{tr} \, \log \, \left[(G^{-1})_{n,m}  +  \tilde{\lambda}_{n,m} \right]~,
\label{detdet}
\end{align}
where the Green function $G_{n,m}$ contains the diagonal component of the $\lambda_{n,m}$:  
\be
\label{propo}
G _{n,m} =\frac1{  2\pi i (n+1/2)+ \lambda_0 } \delta_{n,m}.
\ee 
and $\tilde \lambda$ has Toeplitz form:
\be
\label{tilde}
\tilde\lambda_{n,m}=\left\{\begin{array}{ll}
\lambda_{n-m},&n\ne m\\ 0& n=m \end{array}\right.\,.
\ee
We can expand the logarithm on the right hand side of \eqref{detdet} in a matrix Taylor expansion:
\begin{equation}
\log\det[\partial_\tau + \lambda(\tau) ] = \text{tr} \log G^{-1} + \sum_{n \in \mathbb{Z}^+} \,  \frac{(-1)^{n+1}}{n}\text{tr}  \left(  {G} \cdot \tilde{\lambda} \right)^n~.
\label{detexp}
\end{equation}
The zeroth order piece, using  the standard Euler infinite product formula, can be written as\footnote{The infinite constant is absorbed in the normalization factor $\cal N$  in \eqref{bosonic}.} 
\begin{equation}
\label{det}
 \text{tr} \log G^{-1} = \sum_n \log \left[ 2\pi i (n+1/2)+ \lambda_0 \right] = \log \cosh \frac{\lambda_0}{2}~.
\end{equation}
The first non-trivial contribution in $\tilde\lambda$ to \eqref{detexp} is quadratic and  involves
\begin{equation}
\sum_{n \in \mathbb{Z}}  \frac{1}{\left(2\pi i(n+1/2) + \lambda_0 \right)} \frac{1}{\left(2\pi i(n-m+1/2) + \lambda_0 \right)}  = 0\,, \quad\quad \forall  m \neq 0~.
\end{equation}
The vanishing of the above expression implies that no derivatives of $\lambda(\tau)$  are generated to leading order. It is not hard to check that this behavior continues to higher orders. Thus, the path integral \eqref{bosonic} in a thermal frequency basis becomes
\begin{equation}
\label{bosonicfourier}
Z =\mathcal{N} \, \int \prod_{n \in \mathbb{Z}} d\lambda_n   \cosh^N \frac{\lambda_0}{2} \, e^{-N \gamma \sum_{n \in \mathbb{Z}} \lambda_n \lambda_{-n}}~.
\end{equation}
Performing the integration over the non-zero modes, we arrive at:
\begin{equation} 
Z = \left(  \frac{2^N}{ \int d\lambda_0\,  e^{-  N\gamma \lambda_0^2}} \right) \, {\int  d\lambda_0 \, \cosh^N \frac{\lambda_0  }{2}\, e^{-  N\gamma \lambda_0^2}}~,
\label{lecosh}
\end{equation}
where we have fixed the normalization constant in (\ref{bosonicfourier}) by demanding that in the infinite temperature limit we get the Hilbert space dimension $Z = 2^N$. We see that the partition function can be reduced to a simple integral over $\lambda_0$ (the unique quantity invariant under (\ref{diffeo}) that can be constructed out of $\lambda(\tau)$).
The integral over $\lambda_0$ can be explicitly done. Reinstating $\beta$, we find:
\begin{equation}
Z[\beta] = \sum_{n=0}^N C^N_n e^{\beta(N-2n)^2/(16 N\gamma)}~,
\end{equation}
where $C^N_n$ are the binomial coefficients. From $Z[\beta]$ we can read off the spectrum and degeneracies:
\begin{equation}\label{spectrum}
E_n = -\frac{(N-2n)^2}{16 N\gamma}~, \quad\quad  d_n = C^N_n~.
\end{equation}
Note that the spectrum is symmetric under $n \to (N-n)$.

\subsection{Hilbert space picture}

Can we understand the above result from the point of view of the Hilbert space? Upon quantizing the system, we impose anti-commutation relations $\{\bar{\psi}^I, \psi^J \} = \delta^{IJ}$. If we define an empty state $|0\rangle$ as annihilated by all the operators $\psi^I$, then the full set of states is given by acting with any amount of $\bar{\psi}^I$ on $|0\rangle$. This gives $2^{N}$ states. The Hamiltonian of the system is given by:
\begin{equation}
\hat{H} = -\frac{1}{4N\gamma} \, \left( \bar{\psi}^I \psi^I \right)^2 + \frac{1}{4\gamma} \, \bar{\psi}^I \psi^I - \frac{N }{16\gamma} ~,
\label{H}
\end{equation}
where we have fixed the normal ordering constants such that the spectrum matches that of (\ref{spectrum}). The spectrum is non-positive for $\gamma > 0$. The doubly degenerate ground state has energy $E_g = - N/(16\gamma)$ . Since the operator $\hat{N} = \left( \bar{\psi}^I \psi^I \right)$ commutes with the Hamiltonian, we can organize the spectrum in terms of eigenstates of the $\hat{N}$ operator, which counts the number of $\bar{\psi}^I$ hitting $|0\rangle$. For example $\hat{N} \, \bar{\psi}^1\bar{\psi}^2 |0\rangle = 2\, \bar{\psi}^1\bar{\psi}^2 |0\rangle$.
Thus, states with $n$ $\bar{\psi}$'s acting on $|0\rangle$ have an energy $E_n $ given in (\ref{spectrum}) and a degeneracy given by the binomial coefficients $d_n = C^N_{n}$. As expected, if we sum over all the degeneracies with find: $\sum_n C^N_n = 2^N$. At large $N$ the spectrum peaks sharply about $n = N/2$, with $d_{N/2} \sim 2^{N+1}/\sqrt{2\pi N}$,  consisting of states with $E_{N/2} =0$.

\subsection{Thermodynamic Properties}
\label{thermo}

From the thermal partition function we can study various thermodynamic properties.  At very low temperatures the entropy is $\mathcal{O}(1)$ while the energy goes as $\mathcal{O}(N)$, and the free energy ${ F}=- T\log Z=E -TS$ is dominated by the energy piece.  As we increase the temperature, the energy and entropy contributions begin to compete since the number of states begins to grow exponentially in $N$. More precisely, at large $N$ the degeneracies behave as $d_n \approx C^N_{N/2} \, e^{-(N-2n)^2/(2N)}$. When $d_n=e^{S_n} \approx e^{\beta E_n}$, which at large $N$ gives $\beta \approx 8\gamma$, we expect a transition. Above this temperature, the  entropy   become $\mathcal{O}(N)$ and dominates over the energy. The transition becomes increasingly sharp as we increase $N$. We show a numerical example in figure \ref{thermofig}. Note that in the high temperature phase, the entropy of the system is extensive in the number of degrees of freedom. 

\begin{figure}
\begin{center}
{{\includegraphics[scale=0.43]{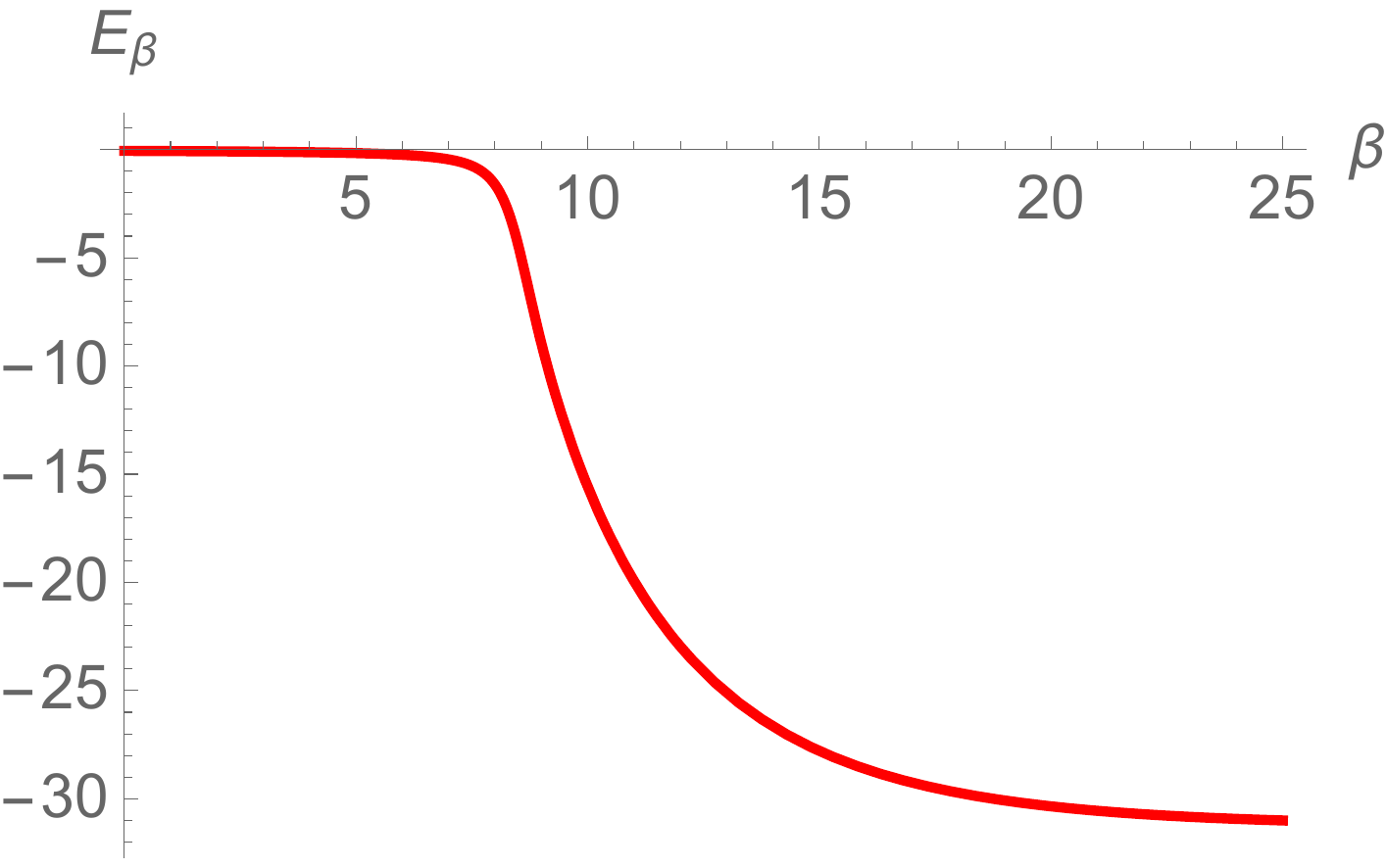}} \quad\quad\quad {\includegraphics[scale=0.43]{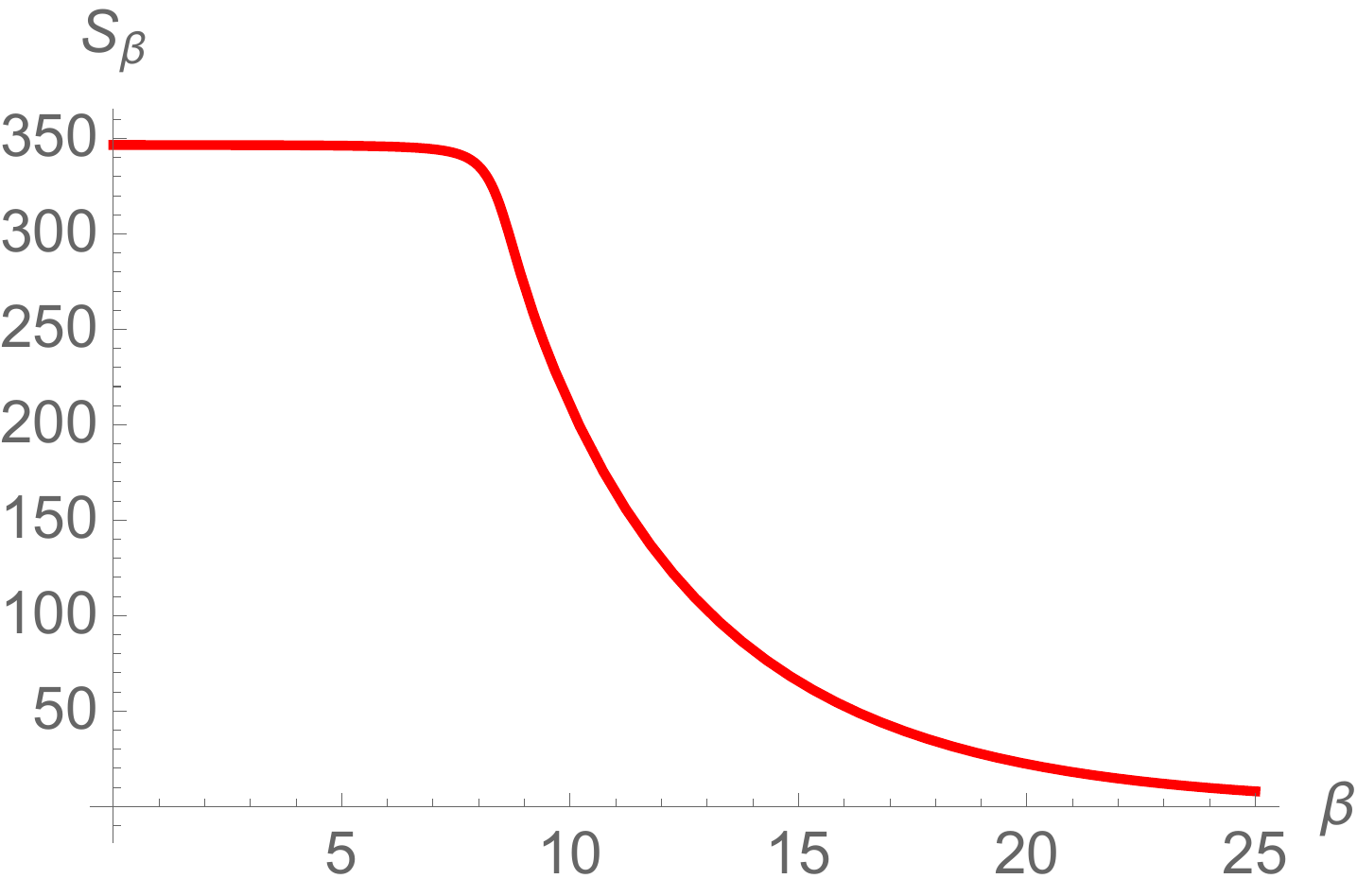} }}
\end{center}
\caption{(a) Energy $\langle E\rangle_\beta=-\partial_\beta\log Z[\beta]$ as a function of $\beta$ (left). (b) Entropy $\langle S\rangle_\beta=(1-\beta\partial_\beta)\log Z[\beta]$  as a function of $\beta$.   We have taken $N=500$ and $\gamma=1$. Notice the transition occurring near $\beta = 8\gamma$.}
\label{thermofig}
\end{figure}

\section{Correlation functions of the vector model}

In this section we discuss the correlation functions of the model in both real and Euclidean time. We do this both in the path integral picture and the corresponding Hilbert space picture.

\subsection{Fermionic Hilbert space picture}

To compute the thermal correlator in the Fermionic Fock space picture, recall that we can organize the Hilbert space in terms of number eigenstates:
\begin{equation}
|n ,  \mathcal{I}_n \rangle = \bar{\psi}^{I_1} \ldots \bar{\psi}^{I_n}|0\rangle~, \quad\quad \langle m ,\mathcal{I} | n,\mathcal{I}' \rangle = \delta_{m,n}~\delta_{\mathcal{I,I}'}.
\end{equation}
The $I_i$ are all different and $\mathcal{I}_n \equiv \{ I_1,I_2,\ldots, I_n\}$ denotes the particular (ordered) collection of creation operators, with $n =0,1,\ldots,N $. The energy and degeneracy of $|n\rangle$ states are given in \eqref{spectrum}.

The {real time} two-point function in the thermal ensemble is explicitly given by:
\begin{equation}\label{Greal}
\langle \psi^A(t) \bar{\psi}^B(0) \rangle_\beta  
		=\frac{\delta^{AB}}{Z[\beta]}\sum_{n=0}^{N-1}C^{N-1}_n e^{-\beta E_n}e^{{-it(E_{n+1}-E_n)}} \,.
\end{equation}
The $(N-1)$ in $C^{N-1}_n$ comes from the reduction in the number of states in $|n,\mathcal{I}_n\rangle$ when hit with the operator $\bar{\psi}^B$. We note that the Green function \eqref{Greal} satisfies $\langle \psi^A(0) \bar{\psi}^B(0) \rangle_\beta =1/2$.

At low temperatures, the two-point function oscillates with a frequency given by the difference in energy of the two lowest lying states. At high temperatures, the dominant contribution to (\ref{Greal}) comes from the most entropic configurations, which have $n\approx N/2$. Moreover, noting that:
\begin{equation}
E_{n+1}-E_n = \frac{N -2 n -1}{4 N\gamma }~,
\label{deltaE}
\end{equation}
we expect to see recurrent patterns in the two-point function for time separations of the order $t\sim N\gamma$. In figure \ref{fig:recplotq} we display a plot of $G_\beta(t)$ exhibiting this behavior. 
\begin{figure}
\begin{center}
{\includegraphics[scale=0.6]{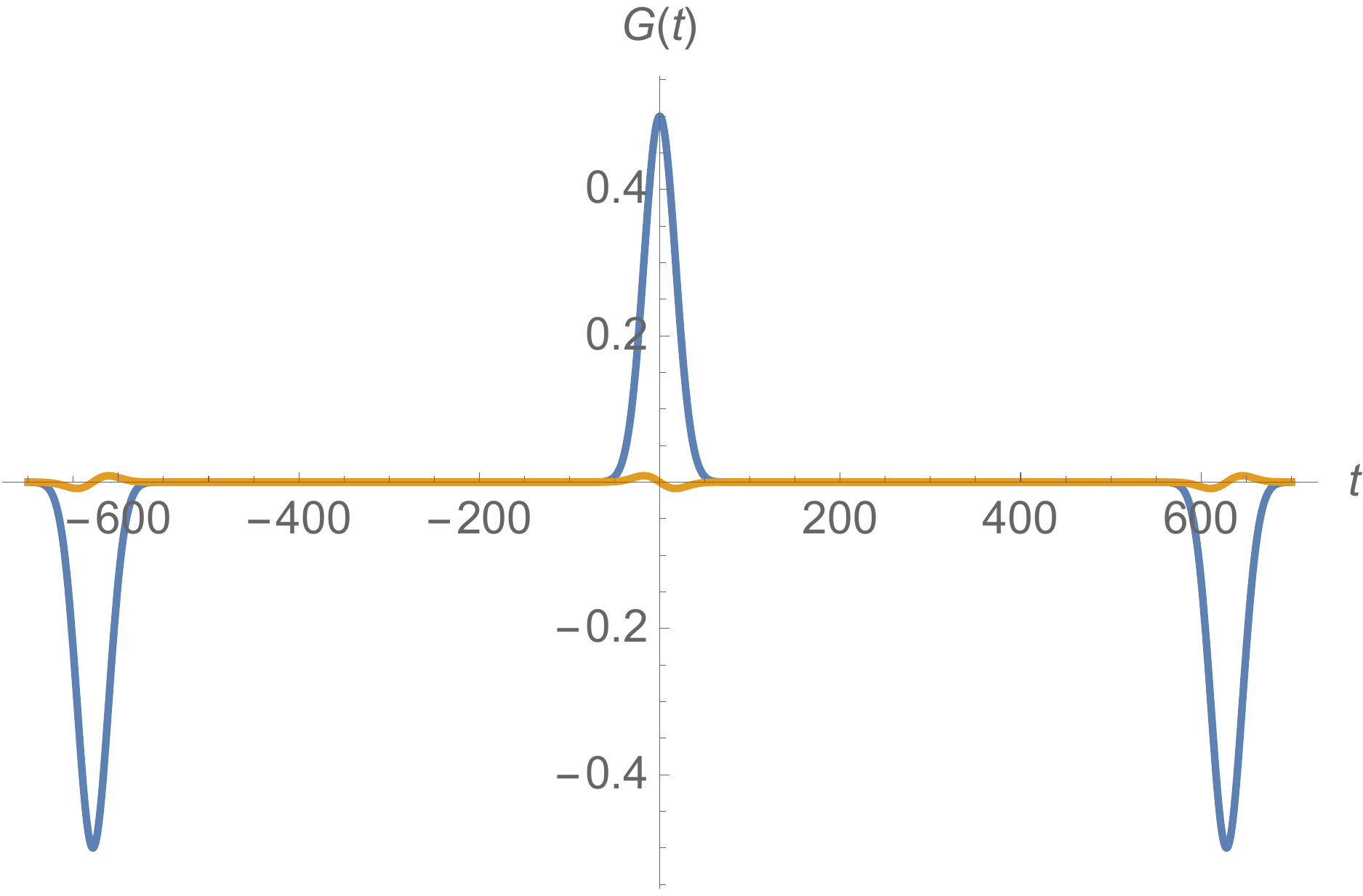}}
\caption{Plot of $G_\beta(t)$ in the high temperature regime for $\gamma/\beta = .5$, and $N=100$. The orange curve is the imaginary part, and the blue curve is the real part. Recurrences occur for $  t\sim 4\pi N\gamma \approx 623$.}\label{fig:recplotq}
\end{center}
\end{figure}

\subsection{Path integral expression}

The generating function for Euclidean fermion correlation functions is given by:
\begin{equation}\label{genZF}
Z[\bar{\xi}_A(\tau),\xi_A(\tau) ] =  \mathcal{N} \int \mathcal{D} \psi^I(\tau) \mathcal{D} \bar\psi^I(\tau)  \mathcal{D} \lambda(\tau) \,  e^{-S_E[\psi^I,\bar\psi^I,\lambda]} e^{-\oint d\tau (\bar{\xi}_A \psi^A+\bar\psi^A \xi_A)}~.
\end{equation}
Integrating out the fermions we get
\begin{equation}\label{genZ}
Z[\bar{\xi}_A(\tau),\xi_A(\tau) ] = \mathcal{N} \int \mathcal{D} \lambda(\tau) \, {\det}^N \left[ \partial_\tau + \lambda(\tau) \right] e^{-N\gamma\oint d\tau \lambda(\tau)^2} e^{\oint d\tau \bar{\xi}_A(\tau) \left( \partial_\tau + \lambda(\tau) \right)^{-1}\xi_A(\tau)}~.
\end{equation}
Consequently, the exact {Euclidean time} two-point function is given by:
\begin{align}\label{prop}
{\bf G}^{AB}_\beta(\tilde\tau) &\equiv \langle \psi^A(\tilde\tau) \bar{\psi}^B(0) \rangle_\beta \nn \\ &=  {\frac{\mathcal{N}}{Z[\beta]}}\int \mathcal{D} \lambda(\tau) \, {\det}^N \left[ \partial_\tau + \lambda(\tau) \right] e^{-N\gamma\oint d\tau \lambda(\tau)^2} \left( \partial_{\tilde\tau} + \lambda(\tilde\tau) \right)^{-1} {\delta^{AB}}~.
\end{align}
The differential operator:
\begin{equation}
{\cal G}(\tau,\tau') = \left( \delta(\tau-\tau') \partial_{\tau'} +  \delta(\tau-\tau')\lambda(\tau') \right)^{-1}~
\end{equation}
obeys the following equation:
\begin{equation}
\int d\tilde{\tau} \, \delta(\tau-\tilde{\tau})\left( \partial_{\tilde{\tau}} + \lambda(\tilde{\tau}) \right) {\cal G}(\tilde{\tau},\tau') = \delta(\tau-\tau')~.
\end{equation}
For $0<{\tau ,\tau'}<\beta$, we find:
\begin{equation}
\mathcal{G}(\tau,\tau') =\left\{\begin{array}{lc} 
e^{-\int_{\tau'}^\tau  d u \, \lambda(u)}\, c_+~, \quad &\text{for}  \quad \tau>\tau'\\
e^{-\int_{\tau'}^\tau  d u \, \lambda(u)}\, c_- ~, ~\quad &\text{for}  \quad \tau<\tau'~.
\end{array}\right.
\end{equation}
The jump in ${\cal G}(\tau,\tau')$ at $\tau=\tau'$ imposes $\lim_{\epsilon\to0} \left[ {\cal G}(\tau'+\epsilon,\tau') - {\cal G}(\tau'-\epsilon,\tau') \right] = 1$, which implies $c_+-c_-=1$. At this point, we must fix the remaining constant. This follows from the fermionic nature of the correlator under a shift in $\beta$, i.e. ${\cal G}(\tau_1+\beta,\tau_2) = -{\cal G}(\tau_1,\tau_2)$. We thus have:
\begin{equation}
c_+ = \left(1 + e^{-\oint  d\tau \lambda(\tau)} \right)^{-1}~.
\end{equation} 
Notice that ${\cal G}(\tau,\tau')$ {is now only invariant under the transformations \eqref{diffeo}  that leave the end points unchanged}. Recalling the interpretation of $\lambda(\tau)$ as an einbein, we can view ${\cal G}(\tau,\tau')$ as a gravitational Wilson line.

Since ${\cal G}(\tau,\tau')$ can be expressed as the exponential of a local integral of $\lambda(\tau)$, we are now in a position to evaluate the path integral. Upon evaluation of the functional determinant, a cancellation occurs between $c_+$ and one of the powers of the $\cosh$ in \eqref{lecosh}, and we can express the path integral as:
\begin{align}
{\bf G}^{AB}_\beta(\tilde\tau) &= \mathcal{N} \frac{ {\delta^{AB}}}{Z[\beta]} \int \mathcal{D}\lambda(\tau)  \sum_{n=0}^{ {N-1}} C^{ {N-1}}_n e^{\oint d\tau \left[ \lambda(\tau) \left(N - 2n \right)/2 -N \gamma \lambda(\tau)^2 \right] }  \, e^{- \int^{\tilde\tau}_0  d \tau \, \lambda(\tau) }\,.
\label{Wilson}
\end{align}
Performing the Gaussian integrals pointwise, and reinstating $\beta$ we find for $0<\tilde\tau < \beta$:
\begin{equation}\label{pathint2}
\langle {\psi}^A(\tilde\tau) \bar{\psi}^{B}(0) \rangle_\beta = \frac{ {\delta^{AB}}}{Z[\beta]} \, \sum_{n=0}^{N-1} C^{ {N-1}}_n  e^{-\beta E_n } e^{-\tilde\tau\left(E_{n+1} - E_{n} \right) }~.
\end{equation}
The normalization constant $\cal N$ has been fixed by imposing ${\bf G}^{AB}_\beta(0)=1/2$. This agrees precisely   with (\ref{Greal}) upon Wick rotating to Euclidean time $t\to-i\tau$. 

Finally, given (\ref{genZ}) we can express all fermionic correlators as expectation values of various collections of Wilson line operators in the bosonic theory. We might view the anti-periodic boundary condition on ${\cal G}(\tau,\tau')$ as being caused by the fermions living at the endpoints of the Wilson lines.

\subsection{Large $N$ approximation}

We would like to see how much of the exact structure previously uncovered is contained in a large $N$ approximation. 
It is convenient to express the partition function in terms of thermal Fourier modes and reinstate $\beta$. From \eqref{det} and \eqref{bosonicfourier} we get:
\be
Z[\beta] = \mathcal{N}  \, \int \prod_{n \in \mathbb{Z}} \,d\lambda_n\,  e^{N\left(\log\cosh\frac{\beta \lambda_0}2 - \beta\gamma\sum_{n \in \mathbb{Z}} \lambda_n \lambda_{-n}\right)}~.
\label{Zfourier}
\ee
The large $N$ saddle point equations for $\lambda(\tau)$ are: \begin{align}
\tanh\frac{\beta \lambda_0}2&=4 \gamma \lambda_0\,,\\
\lambda_n&=0\,,~~~~~\quad~ n\ne0~,
\end{align}
amounting to a constant solution for $\lambda(\tau)$. For low temperature one finds three saddles, which to leading order in the large $\beta$ limit are: $\lambda_0=0$ and $\lambda_0 = \pm1/4\gamma$, the one at the origin being subdominant. As we increase the temperature and reach  $\beta  =8\gamma$  the $\lambda_0\ne0$ saddles coalesce to the origin. Recall that $\beta = 8\gamma$ was the temperature at which the free energy exhibited a thermal transition. For large temperatures, $\beta<8\gamma$, a single saddle is found corresponding to $\lambda(\tau)=0$.

\subsubsection*{\it Low temperatures} 

In the Hilbert space picture, the dominant behavior in the $\beta\to\infty$ limit comes from the vacuum and its first excited state. From \eqref{Greal} we get in the large $N$ limit
\be 
\lim_{\beta \to \infty} \lim_{N\to\infty} {\bf G}_\beta^{AB} (t) = \frac{1}{2} \, e^{- \frac{it}{4\gamma}}, \label{smallTt}
\ee
which shows a single oscillatory behavior in time with frequency given by  the energy difference of the first excited state with respect to the vacuum. In the large $N$ limit this is given by $\Delta E=1/4\gamma$ .

This behavior is recovered, in the path integral description, from the saddles at $\lambda_0 = \pm1/4\gamma$.  The Fourier space expression for \eqref{prop}  is
\be
({\bf G}^{AB}_\beta)_{p,q}=\mathcal{N} \, \frac{\delta^{AB}}{Z[\beta]}\   \int \prod_{n \in \mathbb{Z}} \,d\lambda_n\,  e^{N\left(\log\cosh\frac{ \beta\lambda_0}2 - \beta\gamma\sum_{n \in \mathbb{Z}} \lambda_n \lambda_{-n}\right)} (i\omega_p\delta_{p,q}+\lambda_{p,q})^{-1}~,
\label{GFou}
\ee
with $\omega_p= 2\pi(p+1/2)/\beta $ and $p,q\in\mathbb Z$.  
We can expand the Green function as:
\be
(i\omega_p\delta_{p,q}+\lambda_{p,q})^{-1}=\left((1- G \tilde \lambda+G \tilde \lambda G \tilde \lambda+\ldots)G \right)_{p,q}\,,
\label{pertprop}
\ee
with $G$, and $\tilde\lambda$ defined in (\ref{propo}-\ref{tilde}). Inserting \eqref{pertprop} into \eqref{GFou}, we see that the $\lambda_n$ integrals ($n\ne0$) suppress each term in \eqref{pertprop} order by order by higher powers of $N$. 
To leading order in $N$ we therefore keep  the 1 in \eqref{pertprop}. In the small temperature limit we transform back to $\tau$  replacing $\sum_p\to\beta\int d\omega$ and we also have $\log\cosh\frac{\beta\lambda_0}2\approx \frac{\beta|\lambda_0|}2$. Evaluating \eqref{GFou} at the saddles we get 
\begin{align} 
\lim_{\beta \to \infty} {\bf G} _\beta^{AB}(\tau)&= \, \frac{1}{2}\,  \int d\omega \,e^{i\omega \tau}  \left(  \frac{1}{ i\omega +\frac1{4\gamma}}+ \frac{1}{ i\omega -\frac1{4\gamma}}\right)~\non\\
&=  \frac{1}{2} \, e^{ -\frac{\tau}{4  \gamma }}
\end{align}
which coincides  with \eqref{smallTt} when Wick rotating to real time $\tau\to i t$. Notice that $\tau>0$ ($\tau < 0$) implies that only the $\lambda_0=+1/4\gamma$ ($\lambda_0=-1/4\gamma$) saddle contributes to the $\omega$ contour integral.

\subsubsection*{\it High temperatures} 

Consider first the Hilbert space picture. The high temperature limit of \eqref{Greal}  is dominated by the $n\approx N/2$ terms due to the high degeneracy of states. At large $N$ we can approximate $\sum_n\to\frac N2\int dx$. Using further that $C^{N-1}_n \approx C^{N-1}_{(N-1)/2} \, e^{-(N-1-2n)^2/(2N-2)}$ and defining the variable $x=-(N-(2n+1)))/N$ we get:
\begin{eqnarray}
{\bf G}^{AB}_\beta(t) &\approx& c\, \delta^{AB}  \int dx \, e^{-\frac{x^2 N^2}{(2N-2)}} \, e^{\frac{(N x-1)^2\beta}{16 N \gamma}} e^{\frac{i t x}{4\gamma}}~,\nn \\
&=& \frac{\delta^{AB}}{2} \, e^{-\frac{t^2}{32N \gamma(\gamma-\beta/8)}} e^{-\frac{i t\beta}{32 N \gamma(\gamma-\beta/8)}}~.
\label{hightR}
\end{eqnarray}
Again, the normalization constant $c$ was fixed by demanding that ${\bf G}^{AB}_\beta(0) = 1/2$. In the second line, we have kept only the leading in $N$ terms in the exponent. Interestingly, the correlation functions decay to exponentially small values after a time of order $t \sim \sqrt{N\gamma(\gamma-\beta/8)}$. This is in spite of the correlator not exhibiting an initial exponential decay of the form $e^{-t/\beta}$, characteristic of thermal systems. As mentioned below \eqref{deltaE}, we expect the approximation to fail and find recurrences when all the oscillating factors near $n\approx N/2$ are in phase. This happens for $t\sim 4\pi N\gamma$. So, the recurrences in the vector model occur much more frequently than for a strongly coupled (chaotic) system \cite{Barbon:2014rma}, where they might be separated by exponential in $N$ (or even super-exponential) time scales. These features suggest that the system has a flavor of integrability. The approximate result (\ref{hightR}) agrees well with the exact answer (\ref{Greal}) for times $t\lesssim 4\pi N\gamma$.

We now consider the path integral picture. From \eqref{prop} we have
\begin{align} 
{\bf G}_\beta^{AB}(\tilde\tau) &= \mathcal{N} \,  \delta^{AB} \,  \int \mathcal{D} \lambda(\tau) \, e^{N \left(\log\cosh\frac{\lambda_0}2 \right)}     e^{-N\gamma\oint d\tau \lambda(\tau)^2}  \, e^{- \int^{\tilde\tau}_0  d \tau \, \lambda(\tau) }c_+\, \nn\\
&\approx \mathcal{N} \, \delta^{AB} \, \int da \int \mathcal{D} \lambda(\tau)\,e^{-\frac{8 a^2}{(N-1)}+2 a \lambda_0 + \frac{\lambda_0}{2} } \, e^{-N\gamma\oint d\tau \lambda(\tau)^2} e^{- \int^{\tilde\tau}_0  d \tau \, \lambda(\tau) }\,.
\end{align} 
The auxiliary parameter, $a$, is introduced to make the exponent local in $\tau$. The approximation in the second line corresponds to taking the $\beta\to0$ limit. In this limit $\lambda_0 \ll1$ and we can approximate the $\log\cosh \lambda_0/2$ by a quadratic approximation.  We can now perform the Gaussian integral (\ref{saddlegaussian}) as was done in \eqref{pathint2}, getting
\begin{align}\label{saddlegaussian}
\lim_{N\to \infty}{\bf G}_\beta^{AB}(\tau) = \frac{\delta^{AB}}{2} \, e^{\frac {\tau^2}{32N\gamma(\gamma-\beta/8)}}\, e^{-\frac{\tau \beta}{32 N \gamma(\gamma-\beta/8)}}~,
\end{align}
where again we have fixed the normalization such that ${\bf G}^{AB}_\beta(0) = 1/2$ at $\tau=0$. This coincides with \eqref{hightR} upon Wick rotating. Note that in the large $N$ approximation used above, we have gone being the leading saddle point approximation for which $\lambda(\tau) = 0$, leading to a constant correlation function in $\tau$. The $\tau$-dependence in \eqref{saddlegaussian} arises from the next to leading (Gaussian) correction about the large $N$ saddle point.

As a final note, we should mention that at $\beta = 8 \gamma$, the correlator exhibits a transition between the high temperature decaying correlator and the low temperature oscillatory one. At large $N$ and $\beta=8\gamma$, we must consider the $\log \cosh \lambda_0/2$ term beyond the quadratic approximation. For instance, keeping only the quartic term we are able to approximate the correlator at $\beta = 8\gamma$. In real time, it takes the form:
\begin{equation}\label{8gammaapprox}
\text{Re} \, {\bf G}_{\beta=8\gamma}^{AB}(t) \approx \frac{\delta^{AB}}{2} \left[ {_0}F{_2} \left(\frac{1}{2},\frac{3}{4};\frac{3 t^4}{4N\beta^4}\right)+\frac{\Gamma \left(-\frac{1}{4}\right) \, }{\Gamma \left(\frac{1}{4}\right)}\, \left(\frac{3t^4}{N\beta^4}\right)^{1/2} {_0}F{_2}\left(\frac{5}{4},\frac{3}{2};\frac{3 t^4}{4 N\beta^4}\right)\right]~. 
\end{equation}
The above agrees well with numerics at large $N$, as seen in figure \ref{fig:8gamma}. It falls to small values after $t \sim N^{1/4}\beta$, which is parametrically faster than the high temperature case. 
\begin{figure}
\begin{center}
{\includegraphics[scale=0.6]{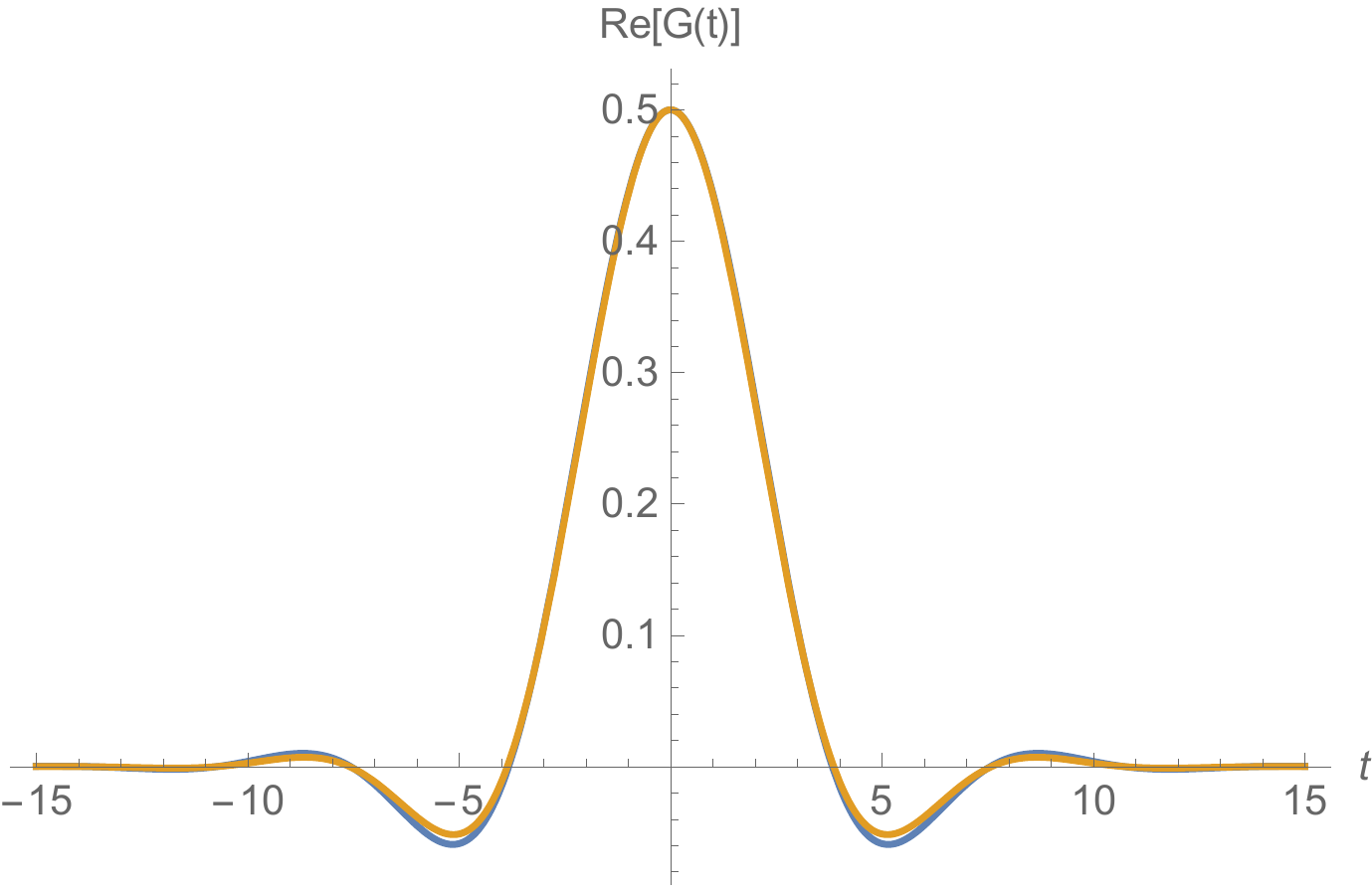}}
\caption{The orange curve is a plot of $\text{Re} \, G_\beta(t)$ for $\beta=8\gamma =1$ and $N=300$. The blue curve is our large $N$ approximation (\ref{8gammaapprox}).}\label{fig:8gamma}
\end{center}
\end{figure}

Our large $N$ analysis is generically not sensitive to the presence of recurrences. Indeed, for large times (\ref{saddlegaussian}) will receive significant corrections from the terms in (\ref{pertprop}) that were discarded. This requires us to go beyond the saddle point approximation. These are responsible in restoring the fine structure of the correlation function.\footnote{Perhaps there is an analogy between this simple model and general observations about correlation functions in black hole backgrounds in the gravity limit \cite{malda2}, where correlation functions also decay to zero at late times.}

\subsection{Alternative view of the low energy sector}

Here, we provide an alternative view to the fermion determinant independence  of the non-zero modes of $\lambda(\tau)$   (\ref{diffeo}) and describe the theory's effective dynamics at low temperature. This approach will be useful when we analyze the fermionic matrix models in what follows. 

The fermionic functional integral:
\begin{equation}
\label{det2}
{\det} [\partial_\tau+\lambda(\tau)] =  \int \mathcal{D} \psi(\tau) \mathcal{D} \bar\psi(\tau)  \,  e^{-\oint d\tau \,  \left({\bar{\psi}}(\tau)  \dot{{\psi}}(\tau) + {\lambda}(\tau)   \bar{\psi}(\tau)  {\psi} (\tau)  \right)}  
\end{equation}
enjoys the following non-compact $U(1)$ gauge symmetry:
\be
\psi(\tau)\to e^{\phi(\tau)}\psi(\tau),~~~\bar\psi(\tau)\to \bar\psi(\tau)e^{-\phi(\tau)},~~~\lambda(\tau)\to\lambda(\tau)-\partial_\tau\phi(\tau)
\label{gt}
\ee
with $\phi(\tau+\beta)= \phi(\tau)$. We can write:
\begin{equation}
\lambda(\tau) = \lambda_0 + \partial_\tau \phi(\tau)~.
\label{gauge}
\end{equation}
The determinant, being gauge invariant, will only depend on $\lambda_0$. A constant piece in $\lambda(\tau)$ cannot be gauged away with a $\phi(\tau)$ periodic in $\tau$.

At low temperatures, we can insert \eqref{gauge} into \eqref{bosonic} and take $\lambda_0$ to be localized at its saddle value $\lambda_0 = 1/4\gamma$. We arrive at an effective low temperature theory:
\begin{equation}
\lim_{\beta \to \infty} Z[\beta] = \mathcal{N} \, e^{\frac{N\beta}{16\gamma}} \int \mathcal{D}' \phi(\tau) \, e^{-N \gamma \int d\tau \left(\partial_\tau \phi (\tau) \right)^2}~,
\label{schw}
\end{equation} 
where $\mathcal{D}'\phi(\tau)$ runs over the space of non-constant periodic functions $\phi(\tau)$. Thus we have at low energy a degree of freedom $\phi(\tau)$, with an ordinary kinetic term. Following the discussion of  \cite{Jensen:2016pah,malda}, this can be viewed as a low energy hydrodynamic mode capturing the fluctuations of the global $U(1)$ charge. 

Fermionic vacuum correlation functions are given by inserting Wilson lines in  the bosonic path integral. Plugging \eqref{gauge} into \eqref{prop} one obtains  a correlation function involving local insertions:
\begin{align}\label{phicorr}
\lim_{\beta \to \infty} \langle \psi^A(\tau_2) \bar{\psi}^B(\tau_1) \rangle = \frac{\delta^{AB} \, e^{-\frac1{4\gamma}(\tau_2-\tau_1)}}{2} \, \frac{\int \mathcal{D}' \phi(\tau)  e^{ \phi(\tau_1)-\phi(\tau_2)} e^{-N \gamma \int d\tau \left(\partial_\tau \phi (\tau) \right)^2}}{ \int \mathcal{D}' \phi(\tau) \, e^{-N \gamma \int d\tau \left(\partial_\tau \phi (\tau) \right)^2} }~.
\end{align}
which evaluates to:
\begin{equation}
\lim_{\beta \to \infty} \langle \psi^A(\tau_2) \bar{\psi}^B(\tau_1) \rangle = \frac{\delta^{AB}}{2} e^{-\left(\tau_2-\tau_1\right)\frac{N-1}{4 N\gamma}}~.
\end{equation}
Once again, we find agreement with the zero temperature correlator of the fermionic theory: 
\be
\lim_{\beta \to \infty}  \langle \psi^A(t_2) \bar{\psi}^B(t_1) \rangle = \frac{\delta^{AB}}{2}  e^{-i(t_2-t_1)\frac{N-1}{4N\gamma}}~.
\label{vacu}
\ee

\section{Fermionic matrix model}

Having understood the fermionic vector model in detail, we now turn to the fermionic matrix model consisting of $N L$ complex fermions $\{ \psi^{i A} , \bar{\psi}^{A i} \}$ with $i = 1,\ldots,N$ and $A=1,\ldots,L$. The indices $i$ and $A$ transform in the bifundamental of a $U(N)\times U(L)$ symmetry. The bosonic variable $\lambda(\tau)$ introduced in \eqref{faction} is replaced by an $N\times N$ Hermitean matrix $M_{ij}(\tau)$, which transforms in the adjoint of the $U(N)$. We introduce the parameter $\alpha \equiv L/N$ for later convenience. The thermal partition function of interest is:
\begin{equation}
Z[\beta] = \int \mathcal{D} \psi^{iA} \mathcal{D} \bar{\psi}^{iA} e^{-\oint d\tau \left(\bar{\psi}^{Ai} \dot{\psi}^{iA} - \frac{1}{4L\gamma   }  \bar{\psi}^{Ai} \psi^{iB} \bar{\psi}^{B j} {\psi}^{j A}\right)}~,
\end{equation}
where as before the fermions are anti-periodic in Euclidean time. It is convenient to consider $\gamma > 0$ such that the quartic term has the opposite sign from the vector case previously studied. At finite $N$, our expressions will be analytic functions of $\gamma$ such that we can analytically continue $\gamma$ over the complex plane. As before, we set $\beta=1$ unless otherwise specified, in the end everything will depend on the dimensionless combination $\gamma/\beta$. 

The corresponding Hilbert space will now consist of $N\times L$ fermionic operators satisfying $\{ \bar{\psi}^{A i} , \psi^{B j}\} = \delta^{ij}\delta^{AB}$, with a $2^{N\times L}$ dimensional Hilbert space. The Hamiltonian governing their interactions will have a quartic term in which the indices are traced in a matrix-like fashion, i.e. $H = - (4L\gamma  )^{-1} \sum_{A,B,i,j} \bar{\psi}^{Ai}\psi^{iB}\bar{\psi}^{Bj}\psi^{jA}$, plus normal ordering terms which will be quadratic. The models are considerably more intricate than their vector counterpart and we will analyze them entirely in terms of the corresponding bosonic path integrals.

\subsection{Effective action}

Prior to integrating out the Grassmann matrix $\psi^{iA}(\tau)$ we have the following Euclidean action:
\begin{equation}
S_E = \oint d\tau \left( \bar{\psi}^{A i } \partial_\tau \psi^{i A} + i \, \bar{\psi}^{A i} M_{ij} \psi^{j A}  +  {L\gamma  \, } M_{ij}M_{ji} \right)~.
\end{equation}
We wish to understand the effective action of $M_{ij}(\tau)$ obtained upon integrating out the $\psi^{iA}(\tau)$. The partition function becomes:
\begin{equation}
Z  = \mathcal{N} \,  \int \mathcal{D} M_{ij}(\tau) \, \det{^L} \left[  \partial_\tau + i \, M_{ij}(\tau) \right] e^{- {L\gamma   } \, \text{tr} \, \oint d\tau M^2}~,
\end{equation}
where $\mathcal{N}$ is a normalization constant which we will fix shortly. 

Note that both the measure over $M_{ij}(\tau)$ and the functional determinant are invariant under the gauge symmetry:
\begin{equation}
M(\tau) \to U(\tau) M(\tau) U^\dag(\tau) -  i \, U(\tau) \partial_\tau U^\dag(\tau)~,
\end{equation}
where $U_{ij}(\tau)$ is a $\tau$-dependent unitary matrix. The above gauge symmetry implies that the determinant is a function of the Polyakov loop:
\begin{equation}
W = \text{tr} \, \mathcal{P} e^{i \oint d\tau M}~,
\end{equation}
where $\mathcal{P}$ denotes path ordering. However, the quadratic in $M_{ij}$ part of the action resembles a mass term of the gauge field and is hence not gauge invariant. Fortunately, one can take advantage of the gauge symmetry in the following sense. We can introduce a Hubbard-Stratanovich field $\Lambda_{ij}(\tau)$ such that:
\begin{equation}
Z = \mathcal{N} \, \int \mathcal{D}\Lambda_{ij}(\tau) \, e^{ - \frac{1}{4L \gamma } \, \text{tr} \oint d\tau \Lambda(\tau)^2} \, \bigg \langle e^{i \text{tr}  \oint d\tau M(\tau) \Lambda(\tau)} \bigg \rangle~.
\end{equation}
Note that the $M_{ij}$ dependent piece of the integrand is simply the generating function of correlation functions of $M_{ij}$, now viewed as a $U(N)$ gauge field in a theory of free fermionic matrices. We can analyze this piece more carefully:
\begin{equation}\label{genfun}
\bigg \langle e^{i \text{tr}  \oint d\tau M(\tau) \Lambda(\tau)} \bigg \rangle \equiv \int \mathcal{D} M_{ij}(\tau)   \, {\det}^L \left[ \partial_\tau + i  M_{ij} \right] \, e^{i \, \text{tr} \oint d\tau \Lambda \, M}~.
\end{equation} 

Under a change of variables, one can write the Hermitean matrix $M_{ij}$ as:
\begin{equation}\label{gaugechoice}
M = U(\tau) \, \mu \, U^\dag(\tau) -  i \,  U(\tau) \partial_\tau U^\dag(\tau)~.
\end{equation} 
Here $\mu = \text{diag}(\mu_1,\ldots,\mu_N)$ is a diagonal matrix with $\tau$-{independent} elements and $U_{ij}(\tau) \in U(N)$. We can thus express (\ref{genfun}) as \cite{minwalla}:
\begin{multline}
\bigg \langle e^{i \text{tr}  \oint d\tau M(\tau) \Lambda(\tau)} \bigg \rangle =  \int [\mathcal{D} U] \,  \prod_{i=1}^N d\mu_i \prod_{i<j} \, \sin^2 \left( \frac{\mu_i-\mu_j}{2} \right)  \times \\  \cos^L \frac{\mu_i}{2}
\, e^{i  \text{tr}  \oint d\tau \, \mu  \,U^\dag \Lambda U} \, e^{-i \text{tr} \oint d\tau \left( i U \dot{U}^\dag \Lambda \right)}~.
\end{multline}
where $[\mathcal{D}U]$ is the $U(N)$ Haar measure.  We now make the transformation $\Lambda = U \tilde{\Lambda} U^\dag$,  leaving the $\Lambda$-measure invariant. The $\mu_i$ integral acquires chemical potentials $\tilde{\lambda}_i \equiv \text{tr}\oint h_i\,\tilde{\Lambda}(\tau)$ for each eigenvalue\footnote{$\tilde\lambda_i$ are the zero modes of the Cartan components of $\tilde \Lambda(\tau)$ with $  h_i$ the Cartan generators of $U(N)$.}, giving rise to a dressed partition function.
All of the $U$ dependence appears in the $\oint d\tau \left( i \dot{U}^\dag U \tilde{\Lambda} \right)$ piece. This piece is independent of the constant part of $\tilde{\Lambda}_{ij}(\tau)$, which we can separate out of the $\text{tr} \, \tilde{\Lambda}^2(\tau)$ term. Performing the $\tilde{\Lambda}_{ij}(\tau)$ integral, we obtain:
\begin{equation}\label{Zi}
Z =\mathcal{N} \, \int [\mathcal{D}U] \, e^{L\gamma   \, \text{tr} \oint  \left( U \dot{U}^\dag \right)^2 } \int \prod_{i=1}^N d\mu_i  \prod_{i<j} \sin^2 \left( \frac{\mu_i-\mu_j}{2} \right) \prod_{i=1}^N \,  \cos^L \frac{\mu_i}{2} \,  e^{-L\gamma  \mu_i^2}~.
\end{equation}
The term $\text{tr} \oint  \left( U \dot{U}^\dag \right)^2$ is analogous to the $\oint d\tau \dot{\phi}(\tau)^2$ in the vector case. It is decoupled from the eigenvalue integral. The normalization constant is given by:
\begin{equation}\label{normalization}
\mathcal{N}^{-1} =  2^{-L} \int [\mathcal{D}U] \, e^{L\gamma   \, \text{tr} \oint  \left( U \dot{U}^\dag \right)^2 } \int \prod_{i=1}^N d\mu_i  \prod_{i<j} \sin^2 \left( \frac{\mu_i-\mu_j}{2} \right) \prod_{i=1}^N  e^{- L\gamma   \mu_i^2}~.
\end{equation}
From the above expression, we can also obtain the case for $\tilde{\gamma} = -\gamma > 0$ by analytic continuation of $\mu_i \to i\mu_i$:
\begin{equation}\label{sinhZ}
\tilde{Z} =\mathcal{Q} \, \int \prod_{i=1}^N d\mu_i  \prod_{i<j} \sinh^2 \left( \frac{\mu_i-\mu_j}{2} \right) \prod_{i=1}^N \,  \cosh^L \frac{\mu_i}{2} \,  e^{-L\tilde{\gamma}   \mu_i^2}\,,
\end{equation}
where now the normalization constant becomes:
\begin{equation}\label{normalization2}
\mathcal{Q}^{-1} =  2^{-L} \int \prod_{i=1}^N d\mu_i  \prod_{i<j} \sinh^2 \left( \frac{\mu_i-\mu_j}{2} \right) \prod_{i=1}^N  e^{-L \tilde{\gamma}  \mu_i^2}~.
\end{equation}
In this case, what was a unitary matrix $U_{ij}$ in (\ref{Zi}) now becomes an element of the group generated by anti-Hermitean matrices.

It is of interest to note that the eigenvalue repulsion is due to a modified version of the usual Vandermonde, that involves the $\sin$ or $\sinh$ of the difference in eigenvalues. This is characteristic of gauge theories at finite temperature \cite{minwalla}.  

It is useful to consider some simple examples, to ensure that the integrals defined above indeed give a positive definite spectrum. Consider the case with $L=1$ and $N=2$, i.e. $2\times 1$ fermionic matrices. Here the Hilbert space is $4 = 2^{2\times 1}$ dimensional. An explicit evaluation of the integral (\ref{sinhZ}) gives:
\begin{equation}
\tilde{Z}[\beta] = e^{\frac{5 \beta }{8 \tilde{\gamma} }} + 3 e^{\frac{\beta }{8 \tilde{\gamma }}}~.
\end{equation}
For $L=2$ and $N=2$ we find:
\begin{equation}
\tilde{Z}[\beta] = 3+ 4 e^{\frac{\beta}{8 \tilde{\gamma} }}+3 e^{\frac{\beta}{4 \tilde{\gamma} }}+4 e^{\frac{3\beta}{8 \tilde{\gamma} }}+e^{\frac{\beta}{2\tilde{\gamma}} }+e^{\frac{3\beta}{4 \tilde{\gamma} }}~.
\end{equation}
The states add up to $16=2^{2 \times 2}$, which is the correct dimension of the Hilbert space. For $L=2$ and $N=3$ we find:
\begin{equation}
\tilde{Z}[\beta] = 4+6 e^{\frac{\beta}{8 \tilde{\gamma} }}+6 e^{\frac{\beta}{4 \tilde{\gamma} }}+10 e^{\frac{3\beta}{8 \tilde{\gamma} }}+8 e^{\frac{\beta}{2 \tilde{\gamma} }}+8 e^{\frac{5\beta}{8 \tilde{\gamma} }}+8 e^{\frac{3\beta}{4 \tilde{\gamma} }}+4 e^{\frac{7\beta}{8 \gamma }}+4 e^{\frac{\beta}{\tilde{\gamma}} }+2 e^{\frac{9\beta}{8 \tilde{\gamma} }}+2 e^{\frac{5\beta}{4 \tilde{\gamma} }}+2 e^{\frac{11\beta}{8 \tilde{\gamma} }}~.
\end{equation}
The states add up to $64=2^{3 \times 2}$, which is the correct dimension of the Hilbert space.

\subsubsection*{\it Summary}

Thus we see that much of the structure of the vector model is carried forward to the matrix model. Instead of an ordinary integral over a single variable, we now have a partition function given by an ordinary matrix integral. Instead of a large $N$ saddle point value for a the single variable, we will find large $N$ eigenvalue distributions. Finally, instead of a single low energy field $\phi(\tau)$, we now have a low energy unitary matrix $U_{ij}(\tau)$. The spectral information about the Hilbert space of the fermionic model is subsumed entirely into the structure of the eigenvalue integral, rather than the low energy fluctuations of $U_{ij}(\tau)$. 

We now explore the thermal phase structure and correlations of the fermionic matrix model.

\section{Thermal phase structure and correlations}

In this section we discuss the thermal phase structure of the matrix model. We will consider the case with $\tilde{\gamma} > 0$. To analyze this we read from \eqref{Zi}  the potential acting on the eigenvalues:  
\begin{equation}\label{eigenvaluepot}
V(\mu_i) =  L \, \sum_{i=1}^N \left( {\tilde{\gamma} }\, \mu_i^2 - \log\cosh\frac{\mu_i}{2} \right)~.
\end{equation}
As before, we fix units where $\beta=1$ and study the partition function as a function of $\alpha \equiv L/N$, which we keep fixed in the large $N$ limit, and $\tilde{\gamma}$.

\subsection{High and low temperature limits}

We can develop a schematic idea of how the eigenvalue distribution should look. The minima of the potential acting on the eigenvalues depends on the temperature. As in the vector model, at low temperatures there are two minima. The eigenvalues will accumulate near these two minima, and will repel each other by the Vandermonde interactions. There will be many such distributions that are solutions. For instance, all $N$ eigenvalues might be located in either minimum or one on the left and all $(N-1)$ others on the right and so on. Eigenvalues in one minimum can thermally tunnel to the other. As we increase the temperature, the double well profile is lost and instead there is a single minimum at the origin. Consequently the eigenvalues will be distributed around a single minimum at high temperatures.

\subsection*{\it {High temperatures}}

In the $\tilde{\gamma} \to \infty$ limit the Gaussian part of the potential acting on the $\mu_i$ dominates. Thus, the partition function receives a large contribution from small values of $\mu_i$ and we can approximate our partition function by expanding the $\cosh\mu_i/2$ in (\ref{eigenvaluepot}) to quadratic order. We obtain the following Gaussian matrix integral:
\begin{equation}
Z \approx \mathcal{Q}  \int d\mu_i \prod_{i<j} { {\sinh^2} } \left( \frac{\mu_i-\mu_j}{2} \right) \, \prod_{i=1}^N e^{-N \alpha  (\tilde{\gamma} - 1/8)  \mu_i^2}~,
\end{equation}
where $\mathcal{Q}$ is defined in (\ref{normalization2}). This matrix integral has appeared in studies of Chern-Simons theory on an $S^3$ \cite{Marino,Tierz:2016zcn}.\footnote{Given the curious connection to Chern-Simons theory and the general relation between matrix models and string theories, it might be interesting to investigate any relation between our fermionic matrix models and topological strings \cite{Marino:2004uf}.} The eigenvalue distribution is connected (single cut) and centered around the origin. Its explicit form is given by:
\begin{equation}\label{marino}
\rho(y) = \frac{1}{\pi \sf t} \tan^{-1} \, \frac{\sqrt{e^{\sf t} - \cosh^2 y/2 }}{\cosh y/2}~,
\end{equation}
with range $y \in   [-2\cosh^{-1}\! e^{\sf t/2},2\cosh^{-1}\! e^{\sf t/2} ]$, where ${\sf t}^{-1} \equiv 2 \alpha (\tilde{\gamma}-1/8)$.

\subsection*{\it Low temperatures}

At low temperatures, i.e. in the $\tilde{\gamma} \to 0$ limit, the eigenvalue potential develops two minima around $\mu_i = \pm1/4\tilde{\gamma}$. Around these, the eigenvalue potential is approximated by:
\begin{equation}
\frac{V_{low}(\mu_i)}{N} \approx  \alpha  \tilde{\gamma} \sum_{i=1}^N \left( \mu_i \pm \frac{1}{4\tilde{\gamma}} \right)^2 - \frac{\alpha}{16 \tilde{\gamma}}~.
\end{equation}
From the result for the eigenvalue distribution \eqref{marino}  we can read  the width of the distribution. 

Consider the case where all eigenvalues are located around one of the minima. We have:
\begin{equation}\label{lowtempmatrix}
\left( \mu \pm \frac{1}{4\tilde{\gamma}} \right) \in \left[-2\cosh^{-1} e^{g/2}, \,2  \cosh^{-1} e^{g/2} \right], \quad\quad g = \frac{1}{2 \alpha \tilde{\gamma}}~.
\end{equation}
In the $\tilde{\gamma} \to 0$ limit we can approximate $\cosh^{-1} e^{g/2} \approx \log 2 + g/2$. Note that for $\alpha > 2/(1-8 \tilde{\gamma} \log 2)$, the eigenvalue distribution has compact support away from $\mu = 0$. For $\alpha < 2/(1-8 \tilde{\gamma} \log 2)$ we cannot have an eigenvalue distribution with compact support away from the origin. More generally however, due to the repulsion of eigenvalues the lowest energy configuration will favor the distribution of eigenvalues evenly among the two minima. Whether the eigenvalue distributions are disconnected will depend on $\alpha$. A small $\alpha$ broadens the potential, enhances the effect of repulsion and connects the two distributions. For parametrically large $\alpha$ the eigenvalues peak sharply about $\mu = \pm 1/4\tilde{\gamma}$. 
\newline\newline
In summary, the global phase structure goes as follows. At large enough temperatures there is a single cut eigenvalue distribution located near the origin. At low temperatures, the eigenvalue distribution senses two minima in the eigenvalue potential, and may be connected (for large $\alpha$) or disconnected (for small $\alpha$). It is interesting to note the similarity of our phase structure to the one studied recently in \cite{sean}. We will present a detailed analysis of the phase structure in future work.

\subsection{Matrix Correlation functions}

Finally, we would like to briefly discuss the correlation functions of $\psi^{iA}(\tau)$. Following the discussion for the vector case, we must invert the differential operator $\mathcal{G}^{-1}_{ij}(\tau,\tau') = \left(\delta(\tau-\tau') \,\partial_\tau + \delta(\tau-\tau')   M_{ij}(\tau)\right)$. Using a parallel argument to the non-matrix case, we find:
\begin{equation}
\mathcal{G}_{ij}(\tau,\tau') =\left\{\begin{array}{lc} 
\mathcal{P} e^{-\int_{\tau'}^\tau d\tau M} \, \left( \mathbb{I}_{N\times N} + \mathcal{P} e^{-\oint M} \right)^{-1}~\quad &\text{for}~  \quad \tau>\tau'~,\\
-\mathcal{P} e^{-\int_{\tau'}^\tau d\tau M} \, \mathcal{P} e^{-\oint M} \,\left( \mathbb{I}_{N\times N} + \mathcal{P} e^{-\oint M} \right)^{-1}~\quad &\text{for}  \quad \tau<\tau'~.
\end{array}\right.
\end{equation}
We see that the inverse operator is naturally expressed in terms of Wilson lines. Note that $\mathcal{G}_{ij}(\tau+\beta,\tau') = -\mathcal{G}_{ij}(\tau,\tau')$. Consequently, at the endpoints of the Wilson lines we have fermionic matrices, rather than vectors. Going to the gauge (\ref{gaugechoice}), we can write for $\tau>\tau'$:
\begin{equation}
\mathcal{P} e^{-\int_{\tau'}^\tau d\tau M} \, \left( \mathbb{I}_{N\times N} + \mathcal{P} e^{-\oint M} \right)^{-1} = U_{ij}(\tau) \,\text{diag}\left(\frac{e^{-(\tau-\tau')\mu_j}}{1+ e^{-\mu_j}}\right) U^\dag_{jk}(\tau')~.
\end{equation}
At large $N$ we calculate:
\begin{equation}
\langle \psi^{iA}(\tau') \bar{\psi}^{Aj}(\tau) \rangle_\beta = \bigg \langle  U_{ij}(\tau)  \frac{e^{-\delta \tau \, \mu_j }}{1 + e^{-\mu_j}}  U^\dag_{jk}(\tau') \bigg \rangle 
\end{equation}
with $\delta\tau \equiv (\tau-\tau')$. Explicitly, we must calculate two pieces. One comes from the $U_{ij}$ sector:
\begin{equation}
\langle U_{ij}(\tau) U^\dag_{jk}(\tau') \rangle_U \equiv \frac{\int [\mathcal{D}U ] e^{L\tilde{\gamma}   \, \text{tr} \oint  \left(U \dot{U}^\dag \right)^2 }  U_{ij}(\tau) U^\dag_{jk}(\tau')}{\int [\mathcal{D}U ] e^{L\tilde{\gamma}   \,\text{tr} \oint  \left( U \dot{U}^\dag \right)^2 }}~.
\end{equation}
The above piece is analogous to the contribution coming from the quadratic $\phi(\tau)$ action, as in (\ref{phicorr}). At large $L\tilde{\gamma}$, The dominant part of the dynamics comes from the eigenvalue piece, which in the large $N$ limit, we can express as an eigenvalue density integral:
\begin{equation}\label{Gtau}
G(\delta \tau)  \equiv  \int dy \, \rho(y) \, \frac{e^{-\delta \tau\, y}}{1 + e^{-y}}~. \end{equation}
For high temperatures, the above integral can be computed numerically using the single cut eigenvalue distribution (\ref{marino}). For Lorentzian times we take $\delta\tau \to i\delta t$. In figure \ref{fig:mar} we show an example $G(\delta t)$.
\begin{figure}
\begin{center}
\includegraphics[width=8cm]{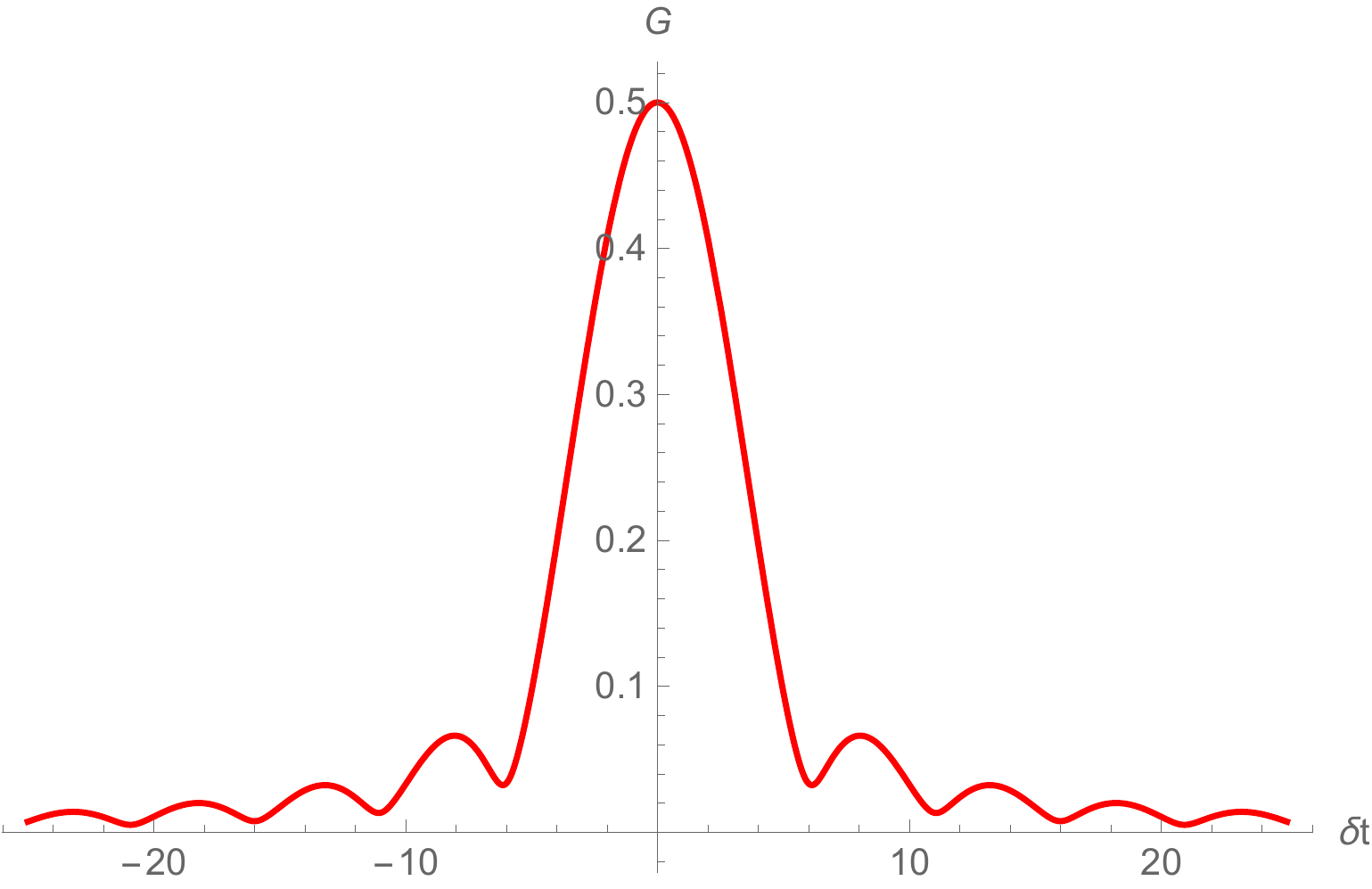}
\end{center}
\caption{Plot of $|G(\delta t)|$ for $g=0.1$.}\label{fig:mar}
\end{figure}
We can give an analytic approximation of $G(\delta\tau)$ at high enough temperatures, or equivalently $\tilde{\gamma}$ large. There, the eigenvalue distribution is close to  Wigner's semi-circle law, so we can approximate \eqref{Gtau} by using $\rho(y) \approx (2\pi {\sf t})^{-1} \, \sqrt{4 {\sf t}-y^2}$. Recalling that ${\sf t}^{-1} = 2\alpha(\tilde{\gamma}-1/8)$ implies that $\sf t$ is a small number at high temperatures. Moreover, the range of $y$ is $\mathcal{O}(\sqrt{\sf t})$ and thus to leading order in small $\sf t$ we find:
\begin{equation}
G(\delta\tau) = \frac{I_1\left({2} \, \sqrt{\sf t} \, \delta \tau \right)}{2 \, \sqrt{\sf t} \, \delta \tau} + \mathcal{O}(\sf t)~,
\end{equation}
where $I_n(z)$ is the modified Bessel function of the first kind. This function agrees well with the numerical result. For $\delta t \gg \sqrt{\alpha \left( \tilde{\gamma} -1/8 \right)}$, we find it decays as $G(\delta t) \sim \delta t^{-3/2}$. This is a significantly faster decay than the vector model at high temperatures, which only experiences a sharp decay after times of order $\delta t \sim 4\sqrt{2 N \gamma(\gamma-1/8)}$. A similar calculation will hold for the low temperature (sub-dominant) saddle where all eigenvalues are gathered around a single minimum and ${\sf t}=1/(2 \alpha \tilde \gamma)$. Higher point functions will be given by computing expectation values of products of Wilson lines. We hope to analyze of the matrix correlators in the  $\beta=8\tilde\gamma$ in future work.

\section*{Acknowledgements}

We would like to acknowledge useful conversations with Tarek Anous, Frederik Denef, Sean Hartnoll, Diego Hofman, Juan Maldacena, Marcos Mari\~no, Dan Roberts, Douglas Stanford and Miguel Tierz. D.A. is funded by the AMIAS and the NSF. GAS would like to thank warm hospitality at IAS during his stay as a Fulbright-CONICET Scholar. G.A.S. acknowledges financial support from projects PICT 2012-0417 ANPCyT, PIP0595/13 CONICET and X648 UNLP. 

\appendix

\section{More general potentials}\label{genpot}

In this appendix, we express the partition function of a vector theory with general potential $V(\bar{\psi}^I \psi^I)$ as a simple integral. One begins by introducing a $\delta$-functional:
\begin{equation}
\delta(\Lambda(\tau) - \bar{\psi}^I(\tau) \psi^I(\tau)) = \int \mathcal{D}\lambda(\tau) e^{i \int d\tau \lambda(\tau) \left( \Lambda(\tau)  - \bar{\psi}^I(\tau) \psi^I(\tau) \right)}~.
\end{equation}
Consequently, we can express the partition function of the fermionic theory as:
\begin{equation}
Z = \int  \mathcal{D}\Lambda(\tau) \mathcal{D}\lambda(\tau) \det{^N} \left( \partial_\tau + i \lambda(\tau) \right) \, e^{-\int d\tau V(\Lambda(\tau))+i \int d\tau \lambda(\tau)  \Lambda(\tau)}
\end{equation}
As in the main text, $ \det \left( \partial_\tau + i \lambda(\tau) \right)$ will only depend on the zero Fourier-mode of $\lambda(\tau)$. Consequently, upon performing the path integral over the non-zero modes of $\lambda(\tau)$, the term $i \int d\tau \lambda(\tau)  \Lambda(\tau)$ in the effective action will produce $\delta$-functions for the $\Lambda_n$ with $n \neq 0$. Hence we will remain with an integral over $\lambda_0$ and $\Lambda_0$. We can write the partition function:
\begin{equation}
Z[\beta] = \mathcal{N} \sum_{n=0}^N C^N_n \int d\lambda_0 \, d\Lambda_0 \,  e^{i \beta(N-2n)\lambda_0/2} e^{i \beta \lambda_0 \Lambda_0} e^{-\beta V(\Lambda_0)}~.
\end{equation}
Integrating over $\lambda_0$ gives us a $\delta$-function $\delta(\Lambda_0 + \beta(N-2n)\lambda_0/2)$, thus allowing us to evaluate the integral in its entirety:
\begin{equation}
Z[\beta] =  \mathcal{N} \sum_{n=0}^N C^N_n e^{-\beta V(\alpha_n)}~, \quad\quad \alpha_n = -\left(N-2n\right)/2~.
\end{equation}
A general $V(\Lambda_0)$ will be an order $N$ polynomial:
\begin{equation}
V(\Lambda_0) = \sum_{i=0}^N \alpha_i \, \Lambda_0^i~.
\end{equation} 

\section{Vector model with $\gamma <  0$}\label{gammaneg}

In this appendix we collect some results on the vector model with $\gamma \equiv -\tilde{\gamma} < 0$. The partition function is now simply:
\begin{equation}
\tilde{Z}[\beta] = \sum_{n=0}^N C^N_n e^{-\frac{\beta\left(N-2n\right)^2}{16\tilde{\gamma}N}}~.
\end{equation}
The bosonic path integral becomes:
\begin{equation}
\tilde{Z}[\beta] = \mathcal{N} \, \int \mathcal{D}\lambda(\tau) \cos^N \frac{\beta\lambda_0}{2} \, e^{-\tilde{\gamma} \, \beta \, N \, \oint d\tau \lambda(\tau)^2}~,
\end{equation}
There is no analogue of the large $N$ low and high temperature phases. The large $N$ saddle point equation is:
\begin{equation}
\tan \frac{\beta  \lambda_0}{2} = -4 \tilde{\gamma}  \lambda_0~.
\end{equation}
The above equation has many solutions, but the dominant saddle lives at $\lambda_0 = 0$. As we lower the temperature, we must include an increasing number of saddles to obtain a good approximation. 

Correlation functions of the fermions are given in the bosonic picture by expectation values of the following non-local operator:
\begin{equation}
\left( \partial_\tau + i \lambda(\tau) \right)^{-1} = e^{-i \int_{\tau_1}^{\tau_2} d\tau \lambda(\tau)} \, \left( 1 + e^{-i\oint d\tau \lambda(\tau) }\right)^{-1}~.
\end{equation}
The correlation function exhibits the Gaussian decay observed for the $\gamma>0$ model discussed in the main text, except that it now does so for all temperatures.

\end{document}